\documentstyle[aps,prbbib,twocolumn,epsf]{revtex}

\begin{document}
\draft
\title{I-V characteristics and differential conductance fluctuations of 
Au nanowires}

\author{H. Mehrez, Alex Wlasenko, Brian Larade, Jeremy Taylor, 
Peter Gr\"utter, and Hong Guo}

\address{Center for the Physics of Materials and Department
of Physics, McGill University, Montreal, PQ, Canada H3A 2T8.}
\maketitle

\begin{abstract}
Electronic transport properties of Au nano-structure are investigated 
using both experimental and theoretical analysis. Experimentally,
stable Au nanowires were created using mechanically controllable 
break junction in air, and simultaneous current-voltage (I-V) and 
differential conductance $\delta I/\delta V$ data were measured.
The atomic device scale structures are mechanically very stable up to
bias voltage $V_b\sim0.6V$ and have a life time of a few $minutes$. 
Facilitated by a shape function data analysis technique which
finger-prints
electronic properties of the atomic device, our data show
clearly differential conductance fluctuations with an amplitude
$>1\%$ at room temperature, and a nonlinear I-V characteristics. 
To understand the transport
features of these atomic scale conductors, we carried out
{\it ab initio} calculations on various Au atomic wires. 
The theoretical results demonstrate that transport properties of these
systems crucially depend on the electronic properties of the
scattering region, the leads, and most importantly the interaction
of the scattering region with the leads. For ideal, clean Au contacts, 
the theoretical  results indicate a linear I-V behavior for bias voltage
$V_b<0.5V$.  When sulfur
impurities exist at the contact junction, nonlinear I-V curves emerge
due to a tunnelling barrier established in the presence of the S atom.
The most striking observation is that even a single S atom can
cause a qualitative change of the I-V curve from linear to nonlinear.
A quantitatively favorable comparison between experimental data
and theoretical results is obtained.  We also report other results 
concerning quantum transport through Au atomic contacts.
\end{abstract}


\section{INTRODUCTION}

Electron transport through atomic nano-contacts has been an active research
area for a decade both experimentally and theoretically. The scientific 
interest of these systems is largely driven by their peculiar electronic
and transport behavior. The atomic nano-contacts are structures 
with low atomic coordination number and, as a result, can behave very 
differently from the bulk counterpart.  From a practical point of view,
understanding the novel electronic and structural properties of the
atomic nano-contacts is an important step towards nano-device fabrication
and characterization. The first set of experiments on nano-contacts 
focused on their zero bias conductance ($G$) using
Scanning Tunnelling Microscopy
(STM)\cite{agrait1993,pascual1993,olesen1994,pascual1995,branbyge1995,agrait1995,smith1995,gai1996,stalder1996,rubio1996,sirvent1996,costa1997a,costa1997b,untiedt1997,cross1998,li1998,jian1999},
Mechanically Controllable Break Junction 
(MCBJ)\cite{krans1993,krans1995,zhou1995,muller1996,scheer1997,scheer1998,yanson1998},
and Relay Contacts (RC)\cite{yasuda1997,hansen1997}. In these experiments a
narrow constriction with a few atoms at the cross section is formed. 
As the electrodes are pulled apart, $G$ is measured and found to
change discontinuously forming plateaus with values close to $n\times G_0$; 
where $n$ is an integer and $G_0=2e^2/h\simeq\frac{1}{12.9K\Omega}$ is 
the conductance quanta. Pioneering experiments\cite{stalder1996,rubio1996} 
clearly showed the correlation between conductance jumps and mechanical 
properties in the nano-contacts. These results confirmed earlier
predictions\cite{tekman1991} that the conductance variations are due to
abrupt changes of nano-structure cross section as a function of
wire elongation. Extensive theoretical investigations on nano-structures
have been published recently to analyze these systems. One major theory 
focus is to calculate the zero bias conductance through a ballistic 
quantum point contact. These calculations start by assuming various
contact\cite{torres1994,torres1996,garcia1996,pascual1997,ruitenbeek1997,lang1997,garcia1997,sanchez1997,wang1997,cuevas1998,bascones1998,martin2000}
geometry, or by using more realistic atomic positions derived from molecular 
dynamics simulations. The potential of the constriction and/or interaction 
Hamiltonian is then
constructed\cite{todorov1993,bratkovsky1995,todorov1996,barnett1997,brandbyge1997a,levy1997,mehrez1997a,mehrez1997b,brandbyge1997b,soresen1998,ohnishi1998,okamoto1999}
from which the zero bias
transport coefficients are evaluated.  While different levels
of approximations were employed in these theoretical analysis, density
functional theory based {\it ab initio} analysis have also been 
reported\cite{lang1997,wang1997,choi1999} which provide self-consistent
calculations of atomic nano-contacts.

While zero bias transport coefficients have received a great deal of
attention, one must go beyond this limit to understand the full 
nonlinear current-voltage (I-V) characteristics of the nano-contacts, 
as this information is essential for the understanding of real
device operation. For example, due to the small cross section of 
these systems, they are exposed to substantial current density 
$\sim 10^8A/cm^2$ which may result in atomic rearrangement. 
It is also expected 
that electron-electron (e-e) interaction is enhanced due to the 
strong lateral confinement possibly leading to Luttinger liquid
behavior for the {\it quasi one dimensional} atomic wires. 
Separating these different effects is experimentally challenging due 
to the many variables which can affect the results. This is probably the 
origin of the existing controversy in explaining the experimentally observed 
non-linear I-V curves of the atomic scale 
wires\cite{costa1997,costa1998,itakura1999,hansen2000}.
From a theoretical point of view, this is also a challenging 
problem: so far only two computationally accurate techniques exist
which can treat systems with {\it open boundaries} out of equilibrium
due to external bias\cite{lang1997,taylor2001,jeremy2}. In the approach of 
Taylor {\it et. al.}\cite{taylor2001,jeremy2}, realistic atomic leads 
can be treated and the problem is solved self consistently within 
LDA. Therefore, the leads, the device (scattering region) and 
their couplings are incorporated without any preconditioned parameters.

To further shed light on the physics of quantum transport at molecular
scale, we report in this paper our investigation on transport properties
of Au nanostructures both experimentally and theoretically. Experimentally,
we created stable Au nanowires using the mechanically controllable 
break junction technique in air, and we simultaneously measure the
I-V curve and the differential conductance $\delta I/\delta V$. We found
that our atomic scale Au nano-contacts are mechanically very stable up to
bias voltage $V_b\sim0.6V$ and have a life time of a few minutes which
is adequate for our measurements. 
As we are interested in features due to electronic degrees of freedom of
the nano-contacts, careful data analysis is needed
because transport data can be
affected by many factors. By defining the {\it shape function},
$S=\frac{\delta I}{\delta V}\frac{V}{I}$, which finger-prints 
electronic properties of the atomic device, our data clearly 
shows differential conductance fluctuations with an amplitude
$>1\%$ at room temperature, and a nonlinear I-V characteristics. 
To understand these transport
features, we carried out {\it ab initio} calculations on various 
Au atomic wires bonded with atomic Au electrodes using
the first principles technique of Ref.\onlinecite{taylor2001,jeremy2}.
Our calculations show that pure and perfect Au nano-contacts
do not give the nonlinear I-V curves as measured in the experiments.
However when Sulfur impurities are present near the wire-electrode
contact region, the nonlinear I-V curves emerge
due to the tunnelling barrier provided by the impurity atoms.
The most striking observation is that even a single S atom can
cause a qualitative change of the I-V curve from linear to nonlinear.
A quantitatively favorable comparison between experimental data
and theory results is then obtained. Our combined theory-experiment
investigation allows us to conclude that transport through Au
atomic wires is strongly affected by the properties of the wire-electrode
contacts.

The rest of the paper is organized as follows.  In the next section,
experimental measurement and results are presented. Section III
presents the theoretical results while Section IV discusses 
the transmission coefficients in more detail. We also discuss and 
compare previous works with ours in section V, followed by a conclusion.

\section{Experimental results}

Atomic scale gold contacts were formed with a mechanically controllable 
break junction in air at room temperature. Once suitably stable atomic 
junctions were formed, a slowly varying bias voltage was applied 
(typically a $0.1 Hz$ triangle wave, $2V_{p-p}$) along with a small
modulation voltage (typically $40 kHz$, $4mV_{rms}$) across the contact 
and a load resistor of $3K\Omega$. Current ($I$) and differential 
conductance ($\delta I/\delta V$) were measured with an I-V
preamplifier and a lock-in amplifier. The experimental set up is shown 
in Fig.\ref{A1}-a. A typical measurement through a stable Au nano-contact 
is presented in Fig.\ref{A1}-b. We show our data for the differential 
conductance and the DC conductance ($G=I/V$). In the inset of 
Fig.\ref{A1}-b, we also display a typical I-V measurement.
This data was taken over the course of a $5$ seconds voltage sweep 
from positive (dark lines) to negative (grey lines) bias voltage ($V$).
Both polarities share a common overall shape, but seem to vary
significantly in the details of their conductance behavior. However, 
for each polarity, one notices that there seem to be similar 
details present in the DC and differential conductance.

An important issue in order to understand these results is 
the separation of effects due to atomic rearrangement in the nano
structure from electronic properties. We address this problem by 
writing, very generally:
\begin{equation}
I(X,V) \equiv g(X) f(X,V)
\label{iv1}
\end{equation}
where the variable $X$ symbolizes the effects of atomic structure, but 
no explicit knowledge of $X$ is required. We define $g(X)$ as the 
unbiased conductance and $f(X,V)$ is the normalized functional form
of the voltage dependence of the current. Therefore, in the 
zero bias limit $V\rightarrow 0$, $f(X,V)=V$. We further define a 
new quantity called the ``shape function" ($S$), 
\begin{equation}
S\equiv \frac{\delta I}{\delta V} \frac{V}{I}\ \ .
\label{S_eq}
\end{equation}
Fig.\ref{A2}-a plots $S$ corresponding to the data of Fig.\ref{A1}-b, 
for both positive and negative bias voltages. The similarity 
between the curves of the shape functions for both polarities, 
including fine details, is in striking contrast to the easily 
distinguishable conductance plots of Fig.\ref{A1}-b. By its
definition (Eq. \ref{S_eq}), $S$ depends only on $f$, the functional 
form of $I(X,V)$, and is independent from the zero-bias conductance
$g(X)$. This fact and the usefulness of the shape function $S$ 
can be seen by considering the following ``Gedanken experiment''. 
Let's assume that we measured $I$ and $\delta I/\delta V$ 
through a variable, ohmic potentiometer as a function of the 
applied voltage. Suppose there were drastic changes in the 
temperature during the measurement, and some troublemaker stochastically 
turned the knob of the variable potentiometer without telling anyone.
Glancing at the measurements of $I(X,V)$ and $\delta I/\delta V$ alone, 
one might wrongfully conclude that the potentiometer was exhibiting 
a nonlinear behavior. However, a plot of $S$ would show that 
$S(V)=1$ (from Eq.\ref{S_eq}), which would allow us to deduce that 
the I-V curve was actually linear and thus in fact ohmic. One would 
also conclude that the origin of the apparently nonlinear
behavior was due to a change in the unbiased conductance $g(X)$,
rather than due to a true nonlinearity in voltage.

Returning to the experimental results shown in Fig.\ref{A2}-a, 
we note the overlap of the two curves of $S$ for positive
and negative bias voltages; this points to a similar
functional form for both bias polarities. Because of this, we
deduce that for this case $f(X,V)$, at most, depends very weakly on 
$X$ (especially at bias voltage $V_b<0.47V$). 
We can therefore write $f(X,V)\simeq f_0(V)\propto e^{\int
\frac{S}{V}dV}$,
and the unbiased conductance 
$g(X)=\frac{I(X,V)}{f_0(V)}$. Hence, as an experimental voltage 
sweep typically takes 5 seconds, time dependent atomic rearrangements 
(changes in $X$) manifest themselves as stochastic variations 
of $g(X)$. 
Other details of this data analysis technique and further discussion 
on the shape function are not within the scope
of this paper and can be found elsewhere\cite{alex2001}.\\

The normalized functional form of the voltage dependence 
of $f_0(V)$ for the measured data is shown in Fig.\ref{A2}-b.
We note that the curves for both polarities appear on the top of 
each other and they are indistinguishable. However,
as expected, the ``trouble maker'' in 
our ``Gedanken'' experiment shows up as fluctuations in 
$g(X)$ shown in Fig.\ref{A2}-c, indicated by the fluctuations 
and by the fact that positive and negative bias give
different traces of $g(X)$.  Note that during the course of our
bias sweep from initial bias voltage to some bias value $V=V_1$, the atomic 
structure has been fluctuating and changed from what
we started with to something unknown. But if we could freeze 
the structure at that moment and re-measure the $I-V$ curve for
the fixed structure, $g(X)$ would be its zero bias conductance.
In other words, Fig. \ref{A2}-c shows a parametric plot of what 
the conductance $g(X)$ would be at zero bias at the point in 
time when this voltage was measured experimentally; hence $g(X)$ could also
be viewed as a function of time in this figure. Our analysis show that
the fluctuations of $g(X)$ which are less than $5\%$ peak-peak, 
are attributed to 
changes in the atomic configuration of the junction. We have also found
that $g(X)$ exhibits no time correlation and its Fourier Transform has a
$1/{\bf f}$ frequency behavior.

In Fig.\ref{A2}-d we plot the {\em normalized} differential conductance, 
$\delta f/\delta V$. Examining the curves in Fig.\ref{A2}-d, 
we can see the same subtle fluctuations in $\delta f/\delta V$ as those
found in $S$ (Fig.\ref{A2}-a). Both plots show broad and fine details
that are uncorrelated to the fluctuations of the unbiased conductance
$g(X)$ (Fig.\ref{A2}-c). These plots reveal that it is the
fluctuations of the $g(X)$ which dominate the fluctuation features in 
the measured $I(X,V)$ and $\delta I/\delta V$.

Over different voltage ranges 
($0.2-0.35V, 0.1-0.35V, 0.1-0.5V$), $\delta f/\delta V$
has a high correlation\cite{rice1995} of $0.99$ between both polarities.
Over these same ranges, $g(X)$ had a weak correlation ($-0.5, 0.25, -0.2$,
respectively) between both polarities which changed drastically depending
on the selected voltage range. We have also calculated the correlations by 
shifting the voltage of {\it one } 
polarity with respect to the other by $\pm 10mV$ (the scale of the fine
details) in increments of $1mV$.
Over all three ranges, the correlation for $\delta f/\delta V$ had a 
local maximum for $0V$ shift. No correlation extrema were found in 
the case of $g(X)$. This quantifies how similar the details in both 
polarities of $\delta f/\delta V$ are to each other, in contrast to 
the more easily distinguishable curves for $g(X)$. We thus conclude
that the ``wiggles" observed in $\frac{\delta f}{\delta V}$ are
electronic in nature and they are not variations in $g(X)$.

The ``wiggles" (magnified in the inset of figure \ref{A2}-d) of 
$\delta f/\delta V$ may actually be much more pronounced than
indicated in these plots. They are smeared out by unavoidable 
experimental constraints. Since we must add a modulation signal 
to make our lock-in measurement of $\delta I/\delta V$,
we end up averaging over a range of $V$. This leads to broadening 
and decrease in amplitude of these fine details\cite{alex2001,eigler}. 
This unavoidable averaging artifact keeps us from making a more 
precise measurement of the shape, amplitude and voltage characteristics of
this fine structure, but the true features should be sharper and more 
pronounced than they are measured to be. 
From our data, we observe ``wiggles" on the voltage scale $<10mV$ and
with amplitude $\sim 1\%$ of the signal. The width of these fluctuations is 
comparable to the modulation voltage, $4mV_{rms} \sim 10mV_{p-p}$; 
hence, it is possible to have features on a voltage 
scale $<10mV$ with amplitude which is orders of magnitude larger than
these $1\%$ fluctuations.

If one were to merely examine our $I(V)$ measurements, similar 
results have been observed in air and at low temperatures
with RC\cite{yasuda1997,itakura1999} and STM 
configurations\cite{hansen2000,hansen2000a}. The fine details exposed 
by the analysis presented above have not been discussed in the 
literature as they tend to be hidden by the fluctuations in the 
unbiased conductance $g(X)$ (see Fig.\ref{A2}-c).
The data presented here represents the behavior of one junction. 
However, we emphasize that the 
general features presented in Fig.\ref{A2} are experimentally 
found to be independent from the zero bias conductance $g(V=0)$ for 
values between 1 - 10 $G_o$. The fine details of $\delta f/\delta V$ 
are reproducible for a given stable junction, but the specific 
details change for different junctions. Although there seems 
to be a voltage scale associated with these details ($<10mV$), 
Fourier analysis does not indicate any strong periodicity. It 
should be noted that we have imposed a selection rule on the junction 
type by studying device configurations that are stable on the time 
scale of minutes. We have also observed junctions that exhibit linear
behavior ($S=1$) as have others\cite{yasuda1997,itakura1999,hansen2000}.
These junctions however are not stable over this long time scale. In comparison to 
their nonlinear counterparts, linear junctions have smaller 
fluctuations in $g(X)$.

In the following we will discuss the physical origin of the observed 
$I(V)$ characteristics. In this aspect we will investigate the electronic 
effects rather than the structural parameters which will be assumed static.
This will allow us to gain valuable insight into the 
voltage dependent conduction properties of nanoscale electrical contacts. 
We will thus compare our modeling to the normalized functional
form of the current, $f_0(V)$, shown in Fig.\ref{A2}-b, rather than the
original I-V curve presented in the inset of Fig.\ref{A1}-b. We will also 
explain the origin of the fluctuations in the normalized differential 
conductance shown in Fig.\ref{A2}-d as well as the effects of temperature 
and modulation signal on their amplitude.

\section{Ab initio analysis of the I-V characteristics}

To provide a theoretical understanding of the experimental data
presented above, we have calculated the I-V characteristics
of Au nano-contacts self consistently by combining the density functional
theory and the Keldysh nonequilibrium Green's functions. The
method is based on the newly developed {\it ab initio} approach to 
treat open electronic systems under finite bias. For technical details 
of this method we refer interested readers to the original 
papers\cite{taylor2001,jeremy2}. Very briefly, our analysis uses
an $s, \ p,\ d $ real space LCAO basis 
set\cite{taylor2001,jeremy2,ordejon1996} and the atomic cores are 
defined by the standard nonlocal norm conserving
pseudopotential.\cite{bachelet1982} The density matrix of the device is constructed
via Keldysh nonequilibrium Green's functions, and the external bias 
$V_b$ provides the electrostatic boundary conditions for the Hartree 
potential which is solved in a three dimensional real space grid. 
Once the density matrix is obtained, the Kohn-Sham effective
potential $V_{eff}({\bf r};V_b)$, which includes contributions from Hartree,
exchange, correlation and the atomic core, is calculated. This process is
iterated until numerical convergence of the self-consistent density matrix is achieved.
In this way, we obtain the bias dependent self-consistent effective
potential $V_{eff}({\bf r};V_b)$, from which we
calculate\cite{taylor2001,jeremy2} the transmission
coefficient $T(E,V_b)\equiv T(E,[V_{eff}({\bf r},V_b)])$, where $E$ 
is the scattering electron energy and $T$ is a function of bias 
$V_b$ through its functional dependence on $V_{eff}({\bf r};V_b)$. 

In our analysis, the scattering states are defined for energy ranges 
between the left and right chemical potentials $\mu_L$ and $\mu_R$, 
respectively. To solve for these states, at a given energy $E$, we 
solve an inverse energy band structure problem\cite{jeremy2}. We 
then group all states as left and right propagating states 
depending on their group velocity. For a scattering state coming 
from the left lead, $\Psi^{K^L_n}$ should start as a right
propagating state $\Phi_{L}^{K^L_n}$ and it gets reflected back 
as a left propagating state $\phi_{L}^{K^L_m}$ with reflection
coefficient $r^{K^L_m,K^L_n}$ in the left lead, and
transmitted into the right lead as a right propagating state
$\phi_{R}^{K^R_m}$ with transmission coefficient $t_{R}^{K^R_m,K^L_n}$. 
In the numerics, the scattering states are represented as a 
linear combination of atomic orbitals inside the device region. 
This allows us to write, for example, a left scattering state as:
\begin{displaymath}
\Psi^{K^{L}_n}=\left \{
\begin{array}{ll}
\Phi_{L}^{k^{L}_n}+\phi_{L}^{K^{L}_m}r^{K^{L}_{m},K^{L}_{n}}
&\textrm{inside left lead}\\
\psi^{K^L_{n}}_d&\textrm{inside device}\\
\phi_{R}^{K^{R}_m}t^{K^{R}_{m},K^{L}_{n}}
&\textrm{inside right lead}
\end{array}
\right.
\end{displaymath}
A scattering state in the right lead can be written in a similar 
fashion. For a symmetric two probe device, the total transmission 
from the left lead is identical to the one from the right 
lead\cite{buttiker1988}. To calculate the total current inside 
the device at a given bias voltage $V_b$ applied to the right 
lead, we use:
\begin{eqnarray}
I(V_b)=& & \frac{2e}{h}\int_{-\infty}^{+\infty}dE \: 
T(E,V_b)[f_L(E,\mu_L=\mu_0)   \nonumber \\
& & -f_R(E,\mu_R=\mu_0+eV_b)]
\label{current}
\end{eqnarray}
where $f_{L(R)}$ is the Fermi function on the left (right) lead 
evaluated at temperature $T= 0K$ unless otherwise stated. In addition 
to predicting the overall transport properties of a device, our 
formalism enables us to study transmission through {\it each} 
incoming Bloch state of the leads separately, therefore allowing us to
separate effects due to the leads and due to the scattering region.
We have used this formalism to calculate I-V characteristics of 
structurally different Au nano-contacts and compared them with the experimental 
results described above.

\subsection{Perfect Au nano-contacts}

In a first attempt to model our experiments, we calculated the I-V 
characteristics of {\it four} Au atoms (=``molecule") in contact with 
Au(100) leads. The structure of the atomic device is illustrated 
in Fig.\ref{fig1}-a. The scattering region is bonded by
two semi-infinite Au leads which extend to electron reservoirs
at $\pm \infty$ where bias voltage is applied and current is collected.
The device scattering region, indicated by D, is described by 
{\it three} Au layers from the left lead, the {\it four} Au 
atoms in a chain, and {\it two} layers of Au from the right lead. 
We have also increased the {\it two} Au layers on the right
side of the chain to {\it four} to ensure that convergence 
is reached with respect to the screening length. 
We note that long and thin gold necks have directly been
observed experimentally\cite{ohnishi1998}.In this 
structure, the registry of the atomic chain with respect to 
lead surface layer can be different. The most common structures, 
which we analyze in this work are: the hollow site, where the atomic 
chain faces the vacant position in the lead layer as shown 
in the upper left inset of Fig.\ref{fig1}-B; and the top site, 
where the atomic chain and an atom from lead surface layer face 
each other as illustrated in the upper right inset of this
figure.

Our calculations show that in all cases charge transfer 
between the Au chain and the leads is not important, being only
$\sim 0.07-0.1$ electrons per atom to the Au chain at different 
bias voltages. This corresponds to less than $1\%$ difference in 
electron population per atom and therefore does not play any 
significant role in the I-V characteristics of Au contacts. 
This is in contrast to a binary atomic system such as carbon chains 
between Al(100) leads\cite{larade2001}. In addition, solving for 
the energy eigenvalues of the {\it four} atom chain gives 
a HOMO-LUMO gap of $0.68eV$, indicating that the molecule 
eigenstates should only have a secondary effect on the transport 
properties. Therefore, the major effect on the I-V characteristics 
is due to the character of Bloch states in the leads and their 
couplings to the molecule at the chain-lead interface.

In Fig.\ref{fig1}-A, we show the results for a system with 
atoms in the hollow site. At small bias $V_b$, we note that current 
is a linear function of $V_b$ with a slope $G\simeq 0.94G_o$. 
The linear function suggests that $T(E,V)=T_0$ with a weak voltage 
dependence. In this regard, our self-consistent calculation gives a result
apparently similar to previous theoretical
work\cite{torres1994,torres1996,garcia1996,pascual1997,ruitenbeek1997,lang1997,garcia1997,sanchez1997,wang1997,cuevas1998,bascones1998,martin2000,todorov1993,bratkovsky1995,todorov1996,barnett1997,brandbyge1997a,levy1997,mehrez1997a,mehrez1997b,brandbyge1997b,soresen1998,ohnishi1998,okamoto1999}
in the low bias regime.
However, we will show later, 
by addressing the origin of this ``perfect linearity'', that
the physical picture of a bias-independent transmission coefficient 
is not valid even for such a simple chain. We also note that
the linear I-V characteristics observed in these systems do 
not agree with our experimental data. 

A major feature of Fig.\ref{fig1}-A is the huge plateau at 
$V_b=0.5V-0.9V$, as well as the fine structures (or sometimes negative 
differential resistance) observed for larger voltages. In the 
lower inset of Fig.\ref{fig1}-B, we show the band structure of 
the Au(100) lead along the $z-$direction (transport direction). 
Even-though at a given energy $E$ there are many Bloch states 
which are potential candidates for transporting current, our 
investigation found that for $E<0.5eV$ there is only {\it one state} 
that is actually conducting (presented by a continuous line in the inset).
Once this state is terminated at $E\approx 0.5eV$, the current is 
saturated resulting in a large plateau until new conducting 
states emerge at higher bias voltages. For $E>0.9V$, transport properties 
are more complex since more states contribute to transmission. Under 
these circumstances, band crossing occurs more frequently and
$T(E,V_b)$ changes over small ranges of bias leading to the small 
structures seen in the I-V characteristics of perfect Au contacts. 
In fact these variations in $T(E,V_b)$ are the origin of the 
fluctuations in the normalized differential conductance shown in 
Fig.\ref{A2}-d. They thus need to be attributed to the
effects of the leads' band structure. In the lower inset of 
Fig.\ref{fig1}-A, we plot the theoretical $\delta I/\delta V$.
The magnitude of these conductance fluctuations is of the order of
$60\%$ at zero temperature which is much larger than the experimental 
finding. However, these fluctuations are reduced to $15\%$ if the current 
is calculated using Eq.\ref{current} at a temperature $T=300K$ 
(the temperature of our experiments). In addition, 
the fluctuations further decrease to $\sim 1\%$ when current is averaged
over the experimental modulation voltage range of $4mV$,
completely consistent with our experimental results of Fig.\ref{A2}-d.

The effect of site registry is studied by placing the end-atom of the
Au chain at the top site of the leads. In this situation the chain atoms 
are facing one atom of the leads' surface layers. This analysis is quite 
important, because it was shown\cite{mehrez1997a,mehrez1997b} that 
atoms at the junction change registry from hollow to top sites
resulting in bundle formation just before the nano structure breaks. The I-V
characteristics of these systems (shown in Fig.\ref{fig1}-B) are similar to the previous 
results, therefore no change is observed in the transport properties
for the top site registry.

To further investigate the huge and peculiar plateau occurring at 
$V_b=0.5V-0.9V$, we have plotted in Fig.\ref{charge} the charge 
density at a given energy $E$, $\rho(E,x,z)=\int dy \rho(E,x,y,z)$.
We see clearly that at zero bias $V_b=0V$, the device property turns from 
a perfect conductor at $E=0eV$ (Fig.\ref{charge}-a) to an insulator at 
$E=0.68eV$ (Fig.\ref{charge}-c) due to the termination of the conducting
state. Here we interpret the charge concentration as the
effective bonding strength, or conductance probability. Applying a bias voltage
to the system drives it out of equilibrium, and at $V_b=0.68V$ and
$E=0.68eV$, the charge in the molecule redistributes, 
but the bonding is still very weak as shown in Fig.\ref{charge}-d, with
some molecular regions having {\it zero charge} and resulting 
in the large plateau
observed in our $I-V$ curve of Fig.\ref{fig1}-A,B.
This effectively demonstrates the importance of both the energy and the 
voltage dependence of the transmission coefficient $T(E,V)$. This point 
will be discussed in more detail in section V.
A further point to notice is the difference of zero bias conductance,
$G\simeq 0.8G_0$ for the top-site device and $G\simeq 0.94G_0$ for the
hollow-site device. This difference is due to a change in the coupling 
between the chain end-atoms and the surface of the leads. To ensure the 
same nearest neighbor separation distance for both cases, we end up with 
{\it four} nearest neighbors for the hollow site registry and only {\it one}
nearest neighbor for the top site registry. Under these circumstances, 
the hollow site has a better coupling to the chain and hence a larger conductance.

A pure and perfect Au nano-contact, as studied in this section, 
shows rich and interesting transport properties. It also gives a good 
understanding of the origin of the observed differential conductance fluctuations 
as due to coupling of the Au-chain to the leads' band structure.
However, it has a linear I-V curve at small $V_b<0.5$V, rather than the
experimentally observed nonlinear I-V characteristics. How can a
nano-contact produce a nonlinear I-V curve such as that of Fig.(\ref{A1}-b) ?
The simplest possibility to observe such a phenomenon is to have a tunnelling 
barrier at the molecule-lead junction whose effect gradually collapses as 
a function of an increasing bias voltage. Indeed, recent experimental findings 
indicate that perfectly linear I-V characteristics were reproducibly found in
gold-gold nano-contacts in ultra high vacuum\cite{hansen2000}, while
nonlinear effects emerge when the experiment was performed in the air. 
This suggests that impurities play an important factor.
Therefore, one has to go one step further and investigate the 
effect of impurity and disorder at the Au junction.

\subsection{Au nano-contacts with S impurity}

To simulate the effect of an impurity at the contact, we have replaced one 
of the Au atoms at the interface layer with a sulphur atom. This is presented 
in the inset of Fig.\ref{fig2}. The choice of sulphur is motivated by the fact 
that in our experimental labs located in down-town Montr\'eal,
sulphur is a non-negligible airborne pollutant (diesel exhausts);
sulphur atoms bond actively with Au. We also note 
that the band-width of sulphur, $\sim 10eV$, is much higher than that of
Au ($\sim 1eV$), thus a tunnelling barrier is expected to be provided by
the presence of S atoms. In this system, charge transfer to the atomic 
chain is still small, and thus inadequate in explaining the experimentally
observed I-V characteristics. We note that the S atom 
suffers from an electron deficiency $\sim 4\%$. Also due to the presence 
of the S atom, 
the coupling of the Bloch states in the leads to the device scattering region 
is quite different as compared to the mono-atomic gold structure. When
the S atom is present, our calculations found that all Bloch states are 
coupled to the scattering region, with the highest transmitting mode 
still being the one corresponding to the conducting mode of the 
mono-atomic gold system. 

The I-V characteristics of the S doped Au nano-contacts is shown in 
Fig.\ref{fig2}. We note that the I-V curve for voltages up to 0.5V is
very similar to the experimental values with nonlinearity onset at 
{\it non zero} bias voltage.
We also note that the huge current plateau of the pure Au 
device has now diminished because more states contribute to electronic transport. 
To compare these results with the experimental measurement, we have
used the normalized functional form of the voltage, $f_0(V)$ shown 
in Fig\ref{A2}-b, and multiplied it by the simulated zero bias conductance ($0.67G_0$).
The result is shown as open circles in Fig.\ref{fig2}. The qualitative and 
quantitative agreement between the theoretical and experimental results 
is rather encouraging.  In fact, this is the first time that 
experimental nonlinear I-V characteristics could be compared so well and
so directly with {\it ab initio} self consistent calculations.
It is also a very surprising result because a single S impurity
can qualitatively alter transport in these nano-contacts from linear
to nonlinear.

The I-V curve in Fig.\ref{fig2} still shows the small features similar
to those found in pure and perfect Au contacts (presented in 
Fig.\ref{fig1}). These fine details of the calculated I-V curve would
result in differential conductance fluctuations similar to the ones shown in the
experimental data of Fig.\ref{A2}-d and the pure Au device of Fig.\ref{fig1}-b. 
However, these fine features as calculated are wider and occur at higher bias voltages 
than the experimentally observed ones. The former can be attributed to the zero temperature 
we used in our calculation. The absence of these fine features at smaller 
voltages is attributed to the small size of the leads used in our theoretical modeling: 
close to $E_F$ there are just a few states so that abrupt variation of 
$T(E,V_b)$ at smaller $V_b$ is less probable, resulting in a smoother 
I-V curve at low $V_b$. 

To understand other possible factors which can affect I-V characteristics of 
nano-contacts we have studied the effect of a larger number of S impurities, 
disorder, and their combined effects. The results of these calculations 
are presented in the following subsection.

\subsection{Contacts with several impurities and disorder}

Including more S impurities at the contact enhances the tunnelling barrier 
and may give rise to a smaller current with a larger nonlinear behavior. 
The result with replacing two Au atoms at the interface by S atoms is
plotted in Fig.\ref{fig3}-A. Indeed, as expected the I-V curve for this system shows a 
larger nonlinear character. The nonlinearity starts 
at $V_b\sim 0.17eV$. It is actually more nonlinear than that of the 
experimental data; this is mainly due to the high concentration 
of S at the interface. Therefore, an experiment which dopes an Au
contact with more S impurity should show an enhancement of 
the I-V nonlinearity. This also gives a possible explanation of why 
experimentally the nonlinear I-V fitting parameters are not 
universal\cite{costa1997,costa1998}: they dramatically
depend on the contact structure as well as the impurity concentration.

Studying the effect of disorder in nano-contacts is another important
problem. To investigate this effect, we have randomized the contact 
layer in the left lead of a pure Au nano-contact, as shown in the inset 
of Fig.\ref{fig3}-B. The distribution of disorder leads to a smaller 
distance between  the contact layer and the Au chain, resulting in a 
better coupling. The current through this device is larger than that 
of the ideal contact and is shown in Fig.\ref{fig3}-B. 
For this device, the slope of the current at zero 
bias is $G\sim 0.915G_o$, and it slightly increases to 
$G\sim 0.92G_o$ at $V_b\sim 0.2V$. The I-V curve shows very weak 
nonlinear characteristics. This suggests that disorder alone may 
create a tunnelling barrier which is overcome through the application of
a bias voltage. However, to observe the effect, there need to be 
conducting states in the scattering region. For our pure Au device, only 
{\it one single state} is conducting and the maximum zero-bias conductance 
is $G=G_o$. Therefore, at $G\sim 0.915G_o$, the conducting channel 
is already open to near its maximum at zero bias, hence it cannot be 
further enhanced in any significant way by applying a bias.
The results in Fig.\ref{fig3}-B show that disorder 
is an important factor that allows more Bloch states in the leads to 
couple with the scattering region and contribute 
to transport properties. This is clearly seen when we notice that 
the huge current plateau observed in Fig.\ref{fig1} essentially vanishes 
in disordered device.

Combining the effect of disorder and S impurity is also crucial. The 
latter enhances the tunnelling barrier and the former may enhance 
the coupling of the device to the leads. We have used the disordered 
structure studied in the last paragraph and replaced one of the Au atoms 
at the contact layer with an S atom, as shown in the insets of
Fig.\ref{fig3}-C. The zero bias conductance for this device is 
$G\sim 0.833G_o$. This number is larger than the one with only a tunnelling 
barrier (ideal contact with S impurity), but is smaller than the one 
with only a disordered contact (which has better coupling). The final 
conductance is due to a competition between both effects. The I-V
curve for this device is presented in Fig.\ref{fig3}-C, 
it shows very weak nonlinear 
I-V characteristics: analysis of 
our data show that the slope reaches $G\sim 0.85G_0$ at $V_b\sim 0.18V$. 
Therefore, there is still a weak nonlinear behavior, but it is
small due to the fast saturation of conducting channel in the device.

From these results we can conclude that the I-V characteristics of an 
atomic junction is a complex phenomenon in which the leads, eigenstates 
of the scattering region, as well as impurity and disorder play major roles. 
In particular, a tunnelling barrier created by impurities can result in 
nonlinear I-V behavior of the nano-device. However, to be able to 
observe this effect, conducting channels in the device need to be 
present, otherwise transmission saturation is reached at small voltages 
and a linear I-V characteristics is seen. Formally, it is the transmission 
coefficient $T(E,V_b)$ that is of crucial importance when analyzing the effect 
of the eigenstates of the leads and the device, as well as the lead-device
coupling. 

In the next section we determine the behavior of 
$T(E,V_b)$ and we will address the following questions: Is the voltage independence of $T(E,V_b)$ an adequate picture? 
Which $T(E,V_b)$ behavior would result in nonlinear I-V characteristics? 
Can the major characteristics of $I(V_b)$ be qualitatively estimated 
from simple arguments or does one always need to perform an extensive 
ab-initio simulations?  

\section{Behavior of transmission coefficient $T(E,V_{b})$}

For all the Au nano-contacts we have investigated theoretically,
$T(E,V_b=0)$ increases as a function of $E$ (for $E<0.2eV$). A typical 
behavior is shown in Fig.\ref{fig4}-a by a dashed line for an
ideal top site, pure and perfect Au device. From this curve, it is 
clear that there is transmission enhancement as a function of
$E$ which should result in a nonlinear I-V curve if 
$T(E,V_b\neq 0)$ behaves in the same way. In the same graph 
we have also plotted $T(E,V_b=0.136V)$. One can observe that the
general energy dependence of $E$ is still the same, namely increasing, 
but there is a global shift of the curve downward. Therefore, an 
increase in transmission coefficient as a function of $E$ is
compensated by its decrease due to increased bias voltage. The total effect 
on the current is to produce a {\it linear} I-V curve as shown in 
Fig.\ref{fig1}-B. We also note that this complete compensation 
between $E$ and $V_b$ is not universal and can be different from one 
system to another. In fact, even for the same device it can behave
differently at different energy ranges. This results in different 
features in the I-V curves such as plateaus, the ``wiggles'', etc.
as we have discussed previously.

A similar analysis is done on a device with a S impurity. 
The results are shown in Fig.\ref{fig4}-b. For this device it is 
clear that the effect of $E$ on transmission at $V_b=0$ is more pronounced. 
This is evidence that tunnelling is an important factor. In addition 
to this, we note that applying a bias voltage to the system causes a 
decrease in the transmission coefficient, but it is not a global decrease. 
In particular, a bias voltage of $V_b=0.136$V actually increases 
transmission at small energies as shown in Fig.\ref{fig4}-b.
Therefore, the combined effect of $E$ and $V_b$ does not cancel and it gives
rise to the nonlinear I-V characteristics as seen in Fig.\ref{fig2}. 
According to this analysis, one can easily predict some aspects of the I-V 
curves just by studying the transmission coefficient $T(E,V_b)$ at two
different bias voltages. Obviously, this helps to predict $I(V_b)$
characteristics with much less computational effort.

In Fig.\ref{fig4}-c,d, we do a similar analysis on a device 
with disorder and one with disorder plus impurity, respectively. Due to 
the conductance channel saturation effect in the scattering region, 
the transmission coefficient has a very weak energy dependence (roughly
constant). This behavior also emerges in the I-V curve which shows 
a very weak nonlinear behavior. We conclude that the roughly linear 
I-V curves of Fig.\ref{fig1}-B and Fig.\ref{fig3}-B,C are due to 
very different origins. In the latter case it is due to the
channel saturation effect in the scattering region, whereas in the 
former it is due to a compensation between the effects of increasing 
energy and bias voltage on $T(E,V_b)$.

\section{Discussions}

We have already shown in the previous sections that 
transport at the molecular level is a complex phenomenon. To understand
these systems, careful experimental work which separates electronic 
effects from structural relaxations, as well as detailed calculations 
which include the effects of the molecule, leads and their coupling 
are required. In this section we discuss and compare our findings 
with previously published theoretical concepts and experimental results.

Experimental work reported by Costa-Kr\"amer 
{\it et. al.}\cite{costa1997,costa1998} have shown clear nonlinear 
I-V characteristics in Au nano-contacts starting at bias $V_b=0.1V$,
and its origin was attributed to strong e-e interactions.
To rule out the impurity effect, the authors\cite{costa1997} 
have used scanning electron microscopy to analyze contacts of diameter $\sim 300nm$. They 
also used energy dispersive X-ray analysis and determined a contamination 
concentration below detection sensitivity. Experimental cleanness 
checks performed for the large nano-contacts ($\sim 300nm$), however, can
not be extrapolated to junctions of few atoms in size due to the
exquisite chemical sensitivity as demonstrated by our model. 
In other experiments\cite{itakura1999} with Au relays, it was observed that
conductance quantization histogram survived for even larger bias, 
($1\times G_o$ peak persists for $V_b\sim1.8V$). These data show 
clearly that in such an experiment nonlinear I-V behavior cannot occur 
at low bias ($V_b\simeq 0.1V$). We believe that these junctions were 
formed between atomically clean gold contacts. These experiments were 
performed by forming and breaking the contacts very quickly ($\sim\mu\,sc$), 
thus removing the impurity atoms from the Au junction even if they
existed\cite{agrait2001}.
Recently, elegant experimental work by Hansen {\it et. al.}\cite{hansen2000} 
found that contaminated Au nanostructure show nonlinear I-V characteristics, 
whereas experiments done with clean tip-sample in UHV show perfect linearity 
for $V_b<0.7V$. The nature of the contamination was not determined. 

A fundamental question that is to be addressed in this section, is
transport through an impurity the most likely physical picture that 
can explain the observed nonlinear I-V characteristics. In the following we
compare and discuss some of the concepts described in the literature 
relevant to this issue.

Free electron models have been used to describe the behavior of 
nanostructures under external bias\cite{pascual1997,martin2000}. In these
systems, depending on the potential profile across the device and
the external bias voltage drop, various I-V characteristics can be 
extracted. In passing, we note that these calculations neglected the 
voltage dependent coupling $T(E,V_b)$ (discussed in Fig.\ref{fig4}). This is obviously a major draw 
back for calculations at the molecular level, as shown by our analysis. 
However, a voltage drop at the contacts is still a reasonable 
approximation. In Fig.\ref{fig5}, we plot the Hartree potential across 
the {\it four} Au-atom chain. It is clearly seen that the potential
drops mostly at the interfaces. However, assuming a uniform potential 
across the constriction is not adequate due to the atomic structure 
and the small variation of charge transfer as a function of bias
voltage.

A self consistent tight binding (TB) model has also been implemented 
to find the conduction dependence of each eigenchannel in Au nano-contact 
as a function of bias voltage\cite{brandbyge1999}. In these calculations, 
the TB parameters are calculated from a bulk system and charge neutrality 
of each atom of the nano-contact is enforced in order to carry out the 
self consistent calculations. Although it is not clear if the TB parameters 
determined from bulk structures are directly transferable to nano-contacts 
with atoms of low coordination number, especially when put under
a bias potential, our {\it ab initio} results show that charge neutrality 
is a valid approximation for Au devices without impurities since charge 
transfer is quite small. In self consistent tight binding models, 
charge neutrality is accomplished by locally adjusting the chemical 
potential. For a zero bias calculation and nano-contacts with 
{\it three} atoms, Ref.\onlinecite{brandbyge1999} shows that 
a local potential $\sim 3eV$ needs to be added to the central atom 
to achieve charge neutrality. This seems to be a large value for
Hamiltonian correction. Therefore, we suggest that for pure 
Au nanostructures, zero bias {\it ab initio} calculations should be 
done to extract the TB parameters (rather than from bulk). These can 
then be used more safely with charge neutrality constraints to deduce 
I-V characteristics of nano-contacts. Since we have seen only small 
effects of the bias voltage on the atomic charge transfer for
a S doped Au structure in our 
{\it ab initio} calculations, we suggest that for this particular 
structure it is possible to deduce the charge distribution and 
the TB parameters from zero bias calculations. 
Using these TB interaction parameters and the zero bias charge at 
each atom as a constraint, the I-V characteristics of these structures 
can then be solved self-consistently. The results are less accurate 
compared to a full {\it ab initio} calculation, but they should give 
better results than using the conventional TB parameters derived from 
bulk systems.

There are other important theoretical calculations going beyond the 
single particle picture. Since the device constriction is narrow, 
e-e interaction can be strong and non-Fermi liquid behavior might have 
to be taken into account. However, it was shown by Maslov and 
Stone\cite{maslov1995} that for these systems the resistance
is due to contact and thus
the results are independent from the Luttinger liquid behavior in the 
constriction. Therefore, in these calculations\cite{maslov1995} 
transmission is found to be identical to non-interacting particles. 
An important, single particle transmission, approximation is incorporated in this
model\cite{maslov1995}. However, at 
non-zero temperature or/and bias voltage, a finite number of particles
are injected into the nano structure resulting in backscattering
effects. These lead to charge accumulation at the interface, 
creating an extra potential in addition to the original constriction
potential. This additional potential is both Hartree ($V_H$) and 
exchange-correlation ($V_{xc}$) in nature, and it is the reason for
the so called resistance dipole\cite{datta_book}. Due to this additional
potential contribution, it was shown\cite{yue1994} that for 1D systems 
the transmission coefficient is renormalized. It was 
further proposed that this charging effect can even 
close a conducting channel that is $90\%$ transmissive\cite{costa1998}.
This channel is gradually opened as $V_b$ is increased for a 
complete transmission at $V_b\sim 0.35V$,
thereby inducing a nonlinear I-V curve. From a theory 
point of view, the picture of charging induced nonlinear 
I-V characteristics should overcome two further difficulties:
that the renormalized transmission depends on an interaction parameter 
$\alpha$ which can not yet be determined for atomic wires; and that 
previous calculations solved a 1D case with no effect of $V_b$ on the 
transmission $T(E,V_b)$. In the rest of this section, we follow the 
interesting idea of the charging effect and analyze it  in more detail to 
understand if this effect, which leads to channel 
closing\cite{costa1997,costa1998} can give rise to nonlinear 
I-V curves of atomic devices. To start, we follow the work
of Yue {\it et. al.}\cite{yue1994} by assuming a strong interaction and
write the renormalized transmission coefficient as\cite{costa1998,yue1994},
\begin{equation}
T^R(E)=\frac{T_0(E/D_0)^{2\alpha}}
{R_0+T_0(E/D_0)^{2\alpha}}
\label{renormT}
\end{equation}
where, $T_0, R_0$ are the transmission and reflection coefficients 
of the non-interacting model such that $T_0+R_0=1$, and $\alpha$ is a 
parameter to describe e-e interaction in the constriction. The parameter 
$D_0$ is the energy range near $E_F$ which contributes to renormalizing 
$T_0$ and is determined\cite{yue1994} by $D_0=\hbar v_F/W$, where $v_F$ 
is the Fermi velocity of the system ($10^8cm/sc$ for Au) and $W$
the width of the nano-constriction\cite{costa1998}, $W\sim10-20\AA$. Therefore, we 
find that for these devices $D_0\sim 1.0eV$. From Eq. \ref{renormT}, 
we compute the current $I(V_b)$ at zero temperature by integrating $T^R(E)$ 
from zero to $eV_b$, assuming no bias dependence of $T^R(E)$.
The conductance is then deduced by $G=I(V_b)/V_b$. We note that due to 
the finite length effect of the constriction\cite{costa1998}, $L\sim 100\AA$, the
renormalization effect is cut off for bias voltages $V_b<V_s$, 
where $V_s=2\pi\hbar v_F/L\sim0.1D_0$. Within this approach and using
the free parameters as specified, we have plotted in Fig.\ref{fig6}-a 
the conductance of a single channel as a function of bias voltage 
$V_b$. Qualitatively, the results show that $G$ increases with $V_b$
due to the channel opening. However, quantitatively they do not give
a complete channel opening at the experimental value of $V_b\sim0.35V$
(as suggested in Ref.\ref{costa1998}), 
if the channel is less than $10\%$ transmissive at $V_b=0$. In fact, 
we found that a bias of $2V$ is needed to overcome the charging potential
barrier, thereby the nonlinearity in I-V curve can only set in at
much larger voltages.
We also note that since the extra charge due to backscattering 
is accumulated at the interface which has a larger cross section, 
we expect that DFT and the local density approximation to 
$V_{xc}$ should work well. Our self-consistent calculations,
with all the charge transfer and re-arrangements accounted for,
have already partially included the backscattering effects.  
Our results show that it is possible to partially close
a channel $\sim 60\%$--but not completely.
Our experimental data which shows zero bias conductance 
of $\sim 2.2G_o$, indicate that at least one channel is $20\%$
transmissive and that channels are {\it only partially closed}, 
consistent with our theoretical work. Therefore, we can 
conclude at this point that the physical picture of electron 
interactions to completely close and open a channel as 
a function of $V_b$, while interesting, cannot explain
our experimental data.

To further address the importance of backscattering effects\cite{yue1994}, 
especially for our devices with realistic atomic leads, 
we have investigated the dependence on the interaction parameter
$\alpha$ in Eq.(\ref{renormT}). Tuning this parameter can dramatically 
change the renormalized transmission coefficient as a function of 
$V_b$ as shown in Fig.\ref{fig6}-b. Therefore we
have a wide range of this parameter for fitting our experimental data. 
However, if we change the Au nano-contact by only a single sulphur atom, 
we should not expect a very different e-e interaction parameter $\alpha$, 
thereby this would predict an I-V curve not too different from the 
one without the sulphur. This suggests that strong interaction alone
would not predict strong sensitivity of the I-V curve dependence on 
small amount of impurities, in contradiction to experimental 
results\cite{itakura1999,hansen2000}.
Transmission renormalization is important in 1D 
scattering problems where there is no charge in the leads. For 
realistic atomic leads, the backscattered charge is a small fraction 
of the original one, and its effect should be well screened.  With
all these considerations, we believe it is unlikely that the dynamically
established charging effect alone is large enough to cause the observed
I-V nonlinearity for Au nano-contacts.

\section{Conclusion}

In this paper, we discuss the electronic transport properties of 
Au nano-contacts from both experimental and theoretical studies. 
Our experimental data analysis enables us to separate electronic 
effects from structural relaxations, allowing a better comparison 
to theoretical modeling of these nano-devices. Our theoretical work
shows that transport properties at the molecular scale need to be 
analyzed at the systems level: leads, molecule and their 
interactions have to be studied simultaneously. The
self-consistently determined transmission coefficient $T(E,V_b)$ 
is shown to vary as a function of $E$ and $V_b$. This gives rise
to differential conductance fluctuations of the order of $1\%$ at 
temperature $300K$ taking into account the experimental averaging 
process. These fluctuations are attributed mainly to the effects 
of the lead band structure. Most striking, however, is the fact
that a single impurity atom at the contact region can alter 
I-V curves qualitatively in these devices: pure and perfect
Au nano-contacts do not give the observed nonlinearity, while
a sulfur doped device does.  Importantly, we
have shown that the measured nonlinear I-V characteristics 
of Au nanowires can be quantitatively modeled by impurity 
effects which create a tunnelling barrier at the nanostructure 
junction.  This effect, among others, points to the vital importance 
for understanding contacts in nano-electronic devices, and, perhaps,
to exploit it for the benefit of device operation.

{\bf Acknowledgments:} We gratefully acknowledge financial 
support from NSERC of Canada and FCAR of Quebec.
H.M. thanks Dr. N. Agra\"{\i}t for a useful discussion 
on their experimental work.

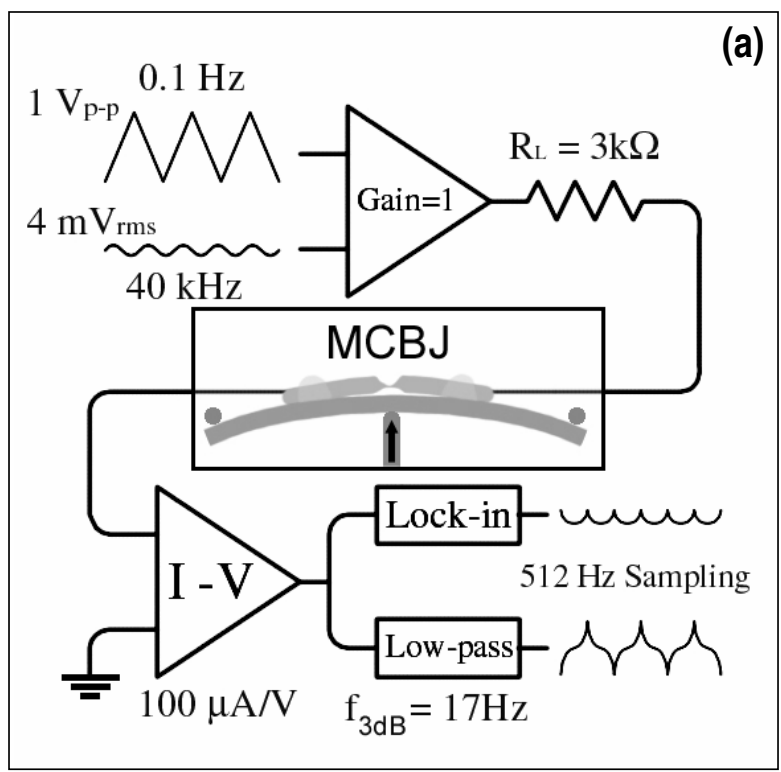
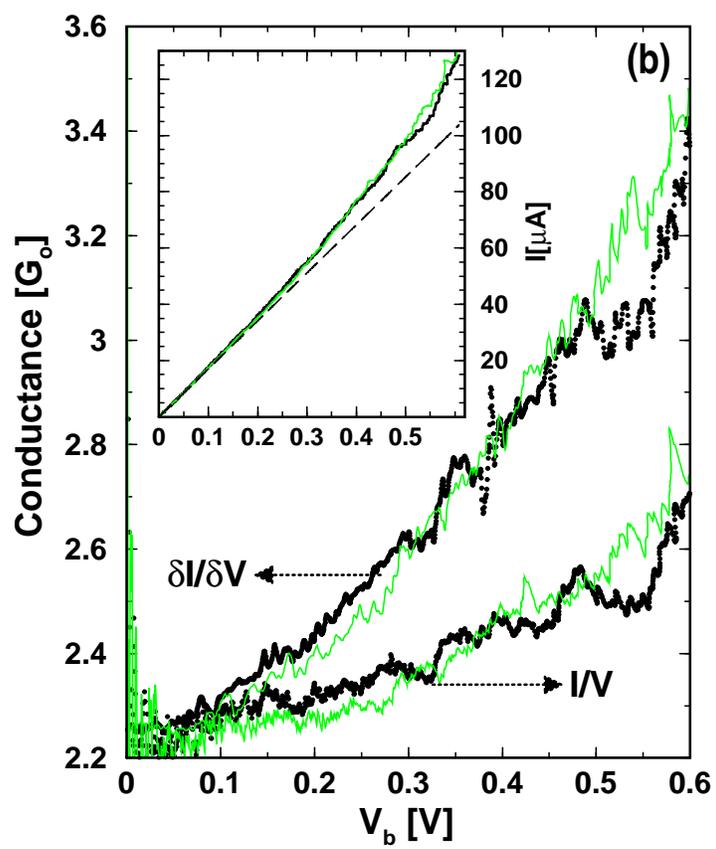
\begin{figure}
\caption{
(a)-Schematic of the experimental set-up with a mechanical controllable break junction (MCBJ) shown at the centre with Au wires
attached. (b)-DC conductance $I/V$ (lower) and differential conductance $\delta I/\delta V$
(upper) plotted vs V; Inset is the I-V curve. 
The dashed line, with slope $\simeq 2.2G_0$, corresponds to linear
behavior and helps to recognize nonlinearity. These 
experimental results are measured over a $5\,second$
voltage sweep from positive (dark) to negative (grey) bias. 
}
\label{A1}
\end{figure}
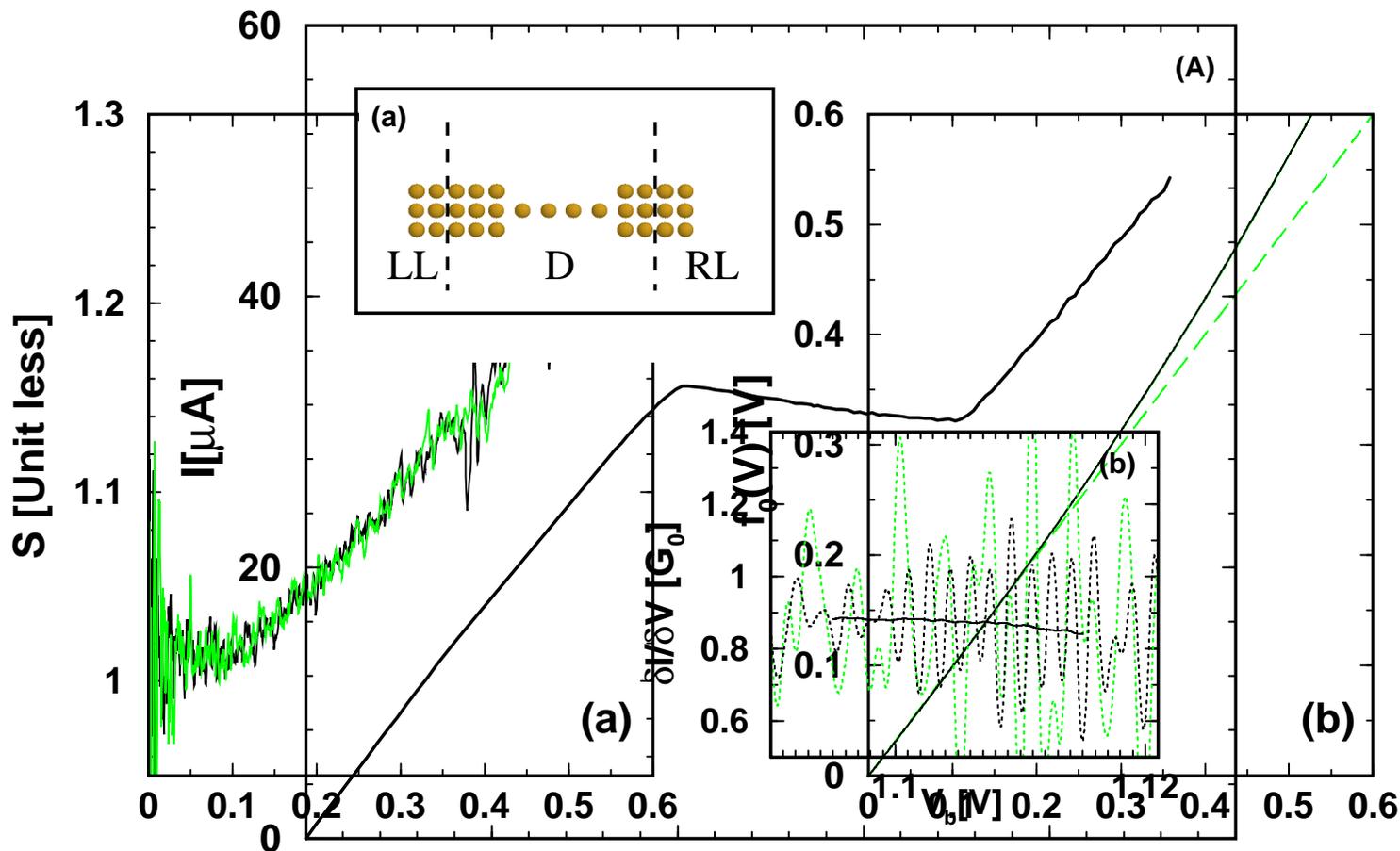
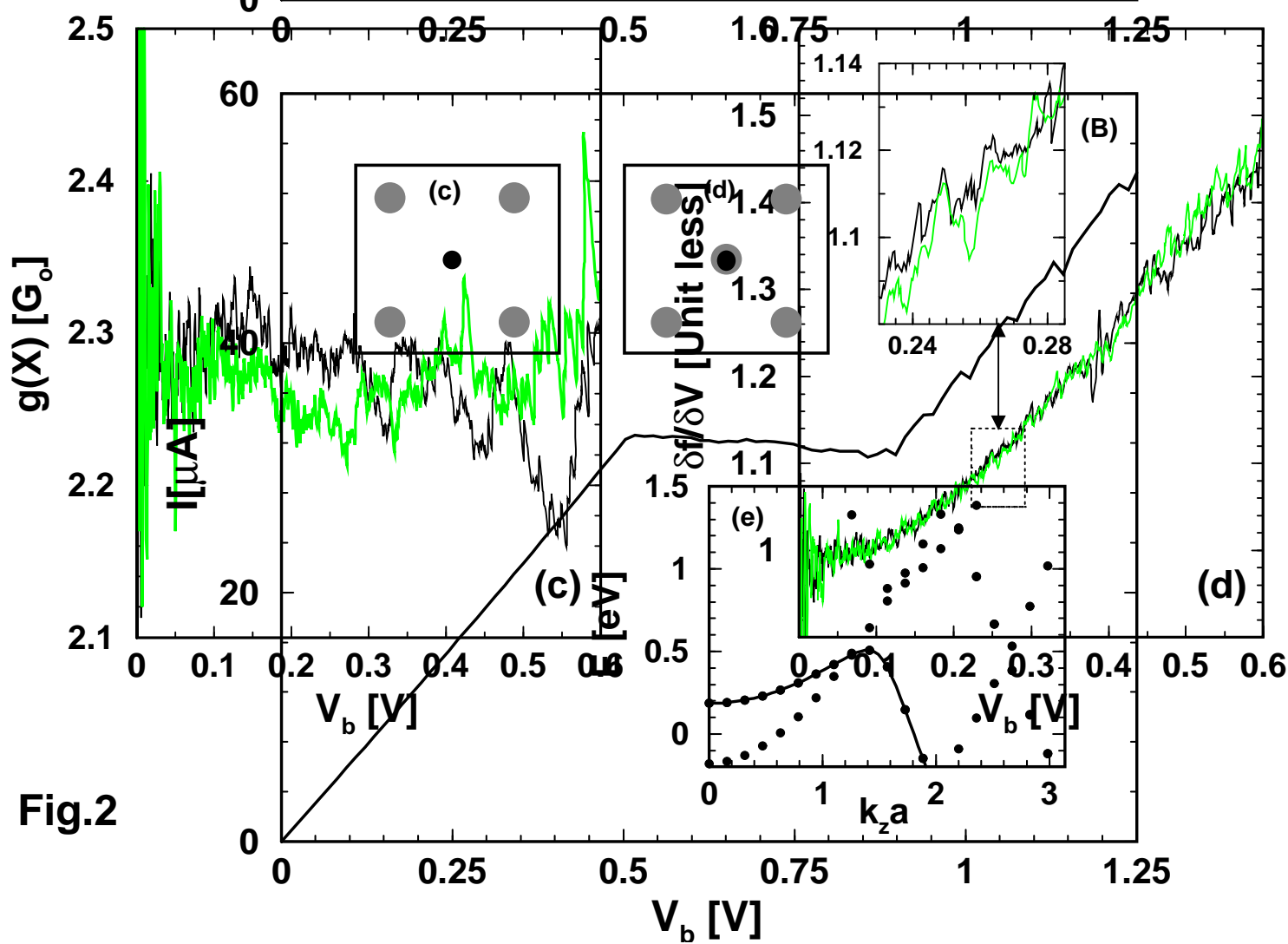
\begin{figure}
\caption{
(a)-The Shape function ($S$,Eq.\ref{S_eq}) vs V calculated from the data shown in Fig.\ref{A1}-b.
(b)-Normalized functional form of the voltage $f_0(V)$ with linear dashed line
shown to help view the onset of non linearity.
(c)- Unbiased Conductance $g(X)$ vs $V$. 
(d)-Normalised conductance, $\delta f/\delta V$ vs $V$; with the inset corresponding
to a magnified section of the same data to show more clearly fine details.
In all graphs, dark (grey) lines
correspond to positive (negative) bias. 
}
\label{A2}
\end{figure}
\begin{figure}
\caption{I-V characteristics of Au contacts with Hollow site registry (A) and Top site
registry (B); with zero-bias conductance $0.94G_0,\; 0.8G_0$ for (A) and 
(B), respectively. The structure of the device is illustrated in (a) where LL, D and RL
correspond to Left Lead, effective Device and Right Lead, respectively. The Hollow site
is shown in (c) and Top site (d) where chain atoms are illustrated by dark circles.
Inset (e) shows the band structure of the lead along the transport direction, the
conducting band is shown by solid line. Inset (b) shows differential conductance
fluctuations of the Hollow site as a grey dotted line, dark dotted line and 
dark continuous line for $T=0K$, $T=300K$ and for $T=300K$ with $4\;mV_{rms}$ 
modulation voltage taken into account; respectively.
}
\label{fig1}
\end{figure} 
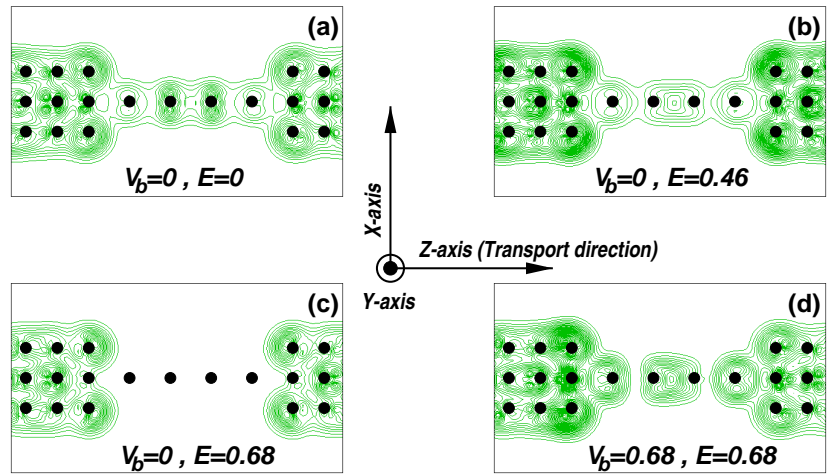
\begin{figure}
\caption{
Contour plots of surface charge density at particular bias voltage $V_b (V)$ and 
Energy $E (eV)$ indicated on each figure. We use the same scale for all graphs to 
compare the conductance probability for each configuration. Dark circles correspond
to atom positions along the constriction for guidance.}
\label{charge}
\end{figure}
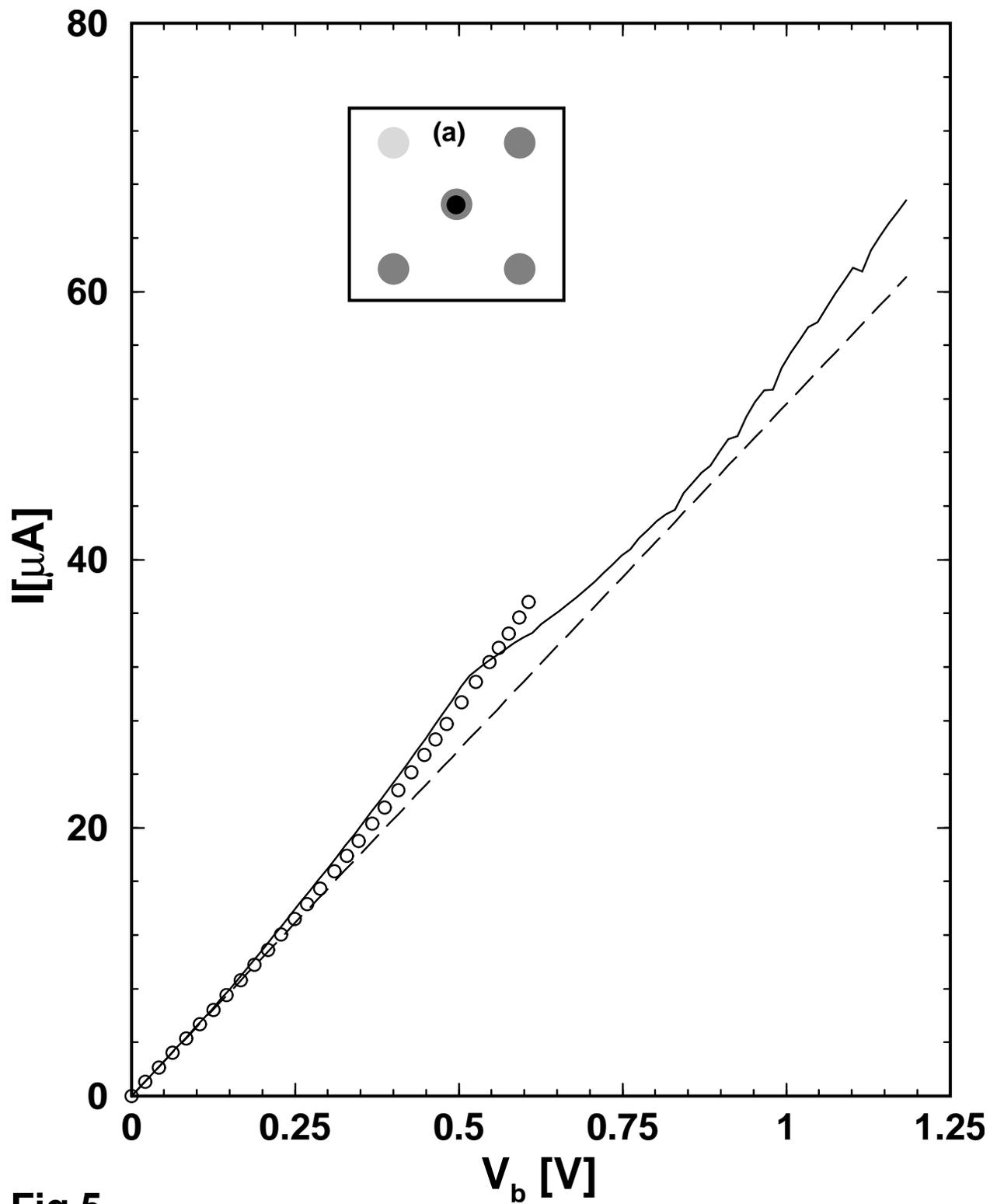
\begin{figure}
\caption{I-V characteristics of Au contact doped with S impurity (Solid line). The dashed
line has a slope $\simeq 0.67G_0$ and is shown to 
help view the onset of non linearity; Circles
correspond to the experimental results obtained from $f_0(V)$ (Fig.\ref{A2}-b) and multiplied by the
zero bias conductance, $0.67G_0$. The
inset (a), shows the atoms registry, where dark circles show the atomic chain, grey
ones are Au atoms in the leads' surface and light grey is the S atom.
}
\label{fig2}
\end{figure}
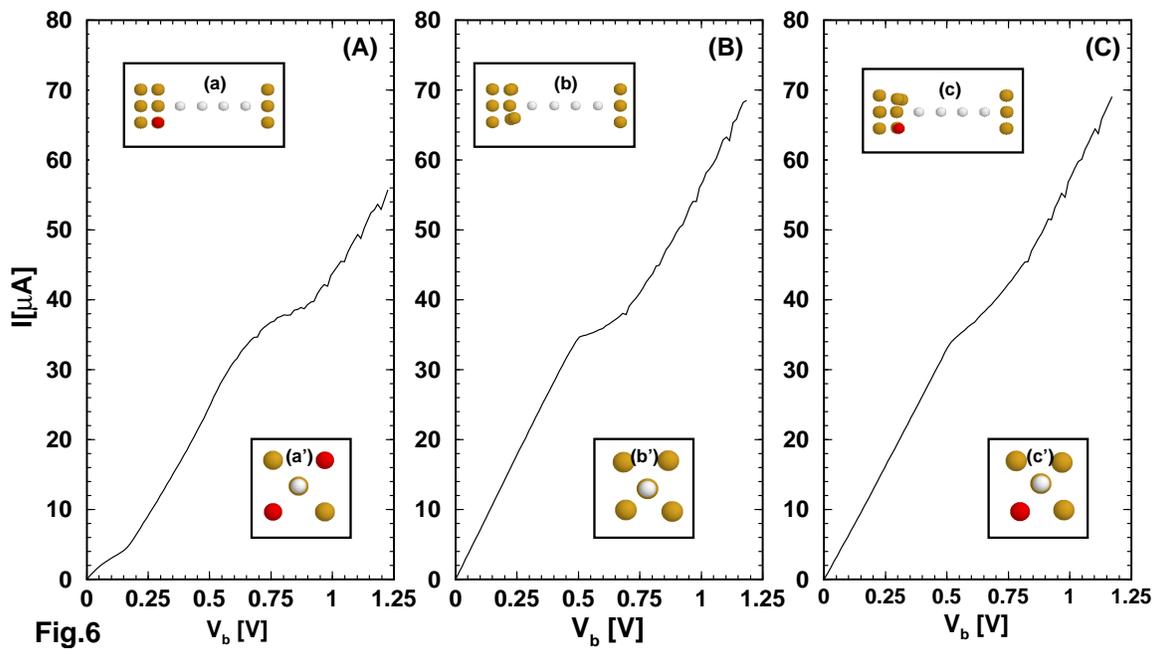
\begin{figure}
\caption{I-V characteristics of Au contacts with 2 S impurity (A), disordered
interface (B) and disordered interface with 1 S impurity (C). The insets illustrate the
corresponding structures, with S represented as a dark circle; Au atoms in the
lead and the chain are shown with grey and light grey circles, respectively.
(A), (B) and (C) have zero-bias conductance of $0.55G_0,\; 0.915G_0$
and $0.833G_0$, respectively.
}
\label{fig3}
\end{figure}
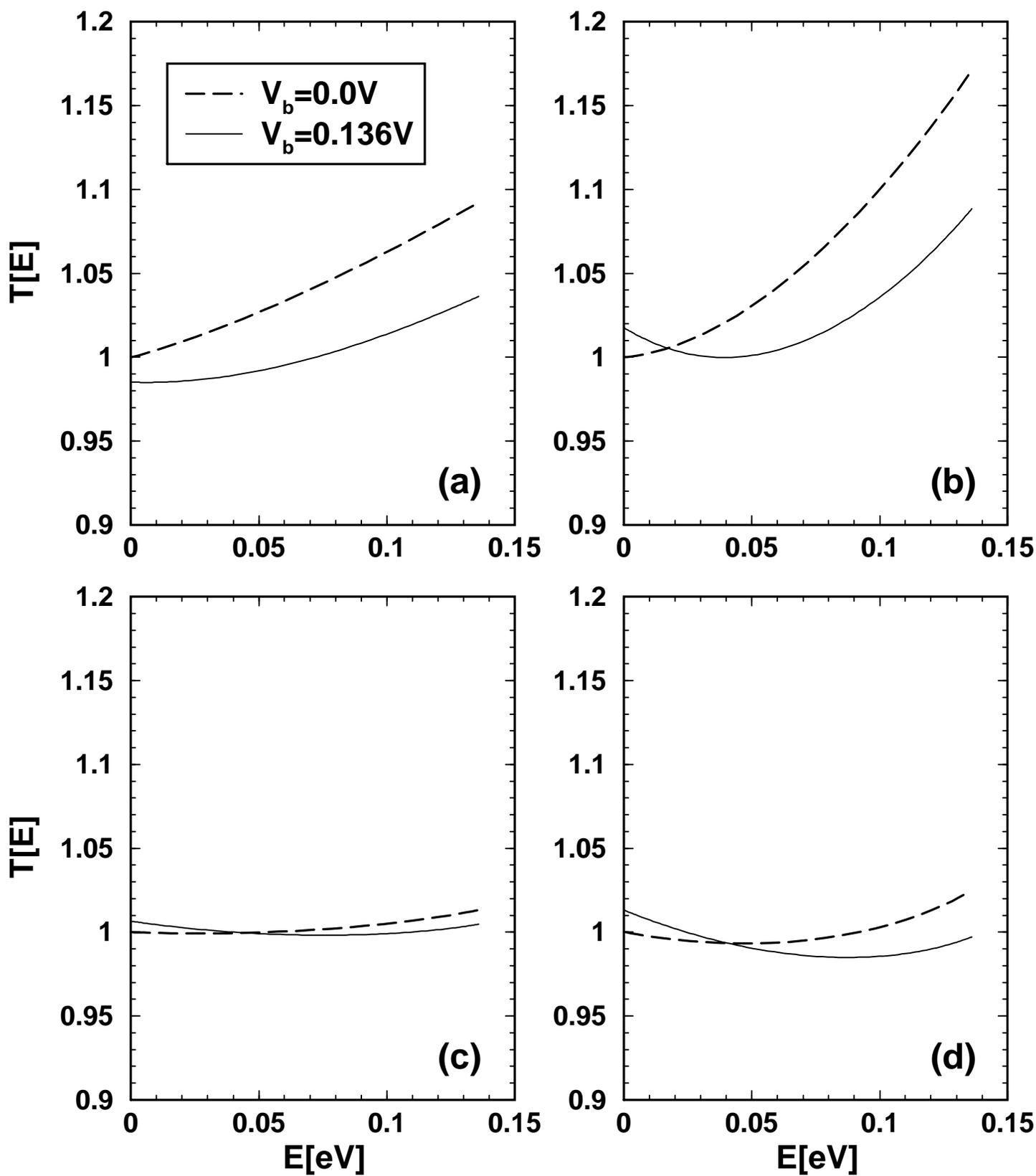
\begin{figure}
\caption{ Scaled transmission coefficient $T(E)$ at $V_b=0$ (dashed line) and $V_b=0.136V$ 
(solid line) for ideal contact structure at the top site (a), 1 S impurity structure (b), disordered
interface (c) and disordered interface with 1 S impurity (d). 
The scaling factor is $T(E=0,V_b=0)$.
}
\label{fig4}
\end{figure}
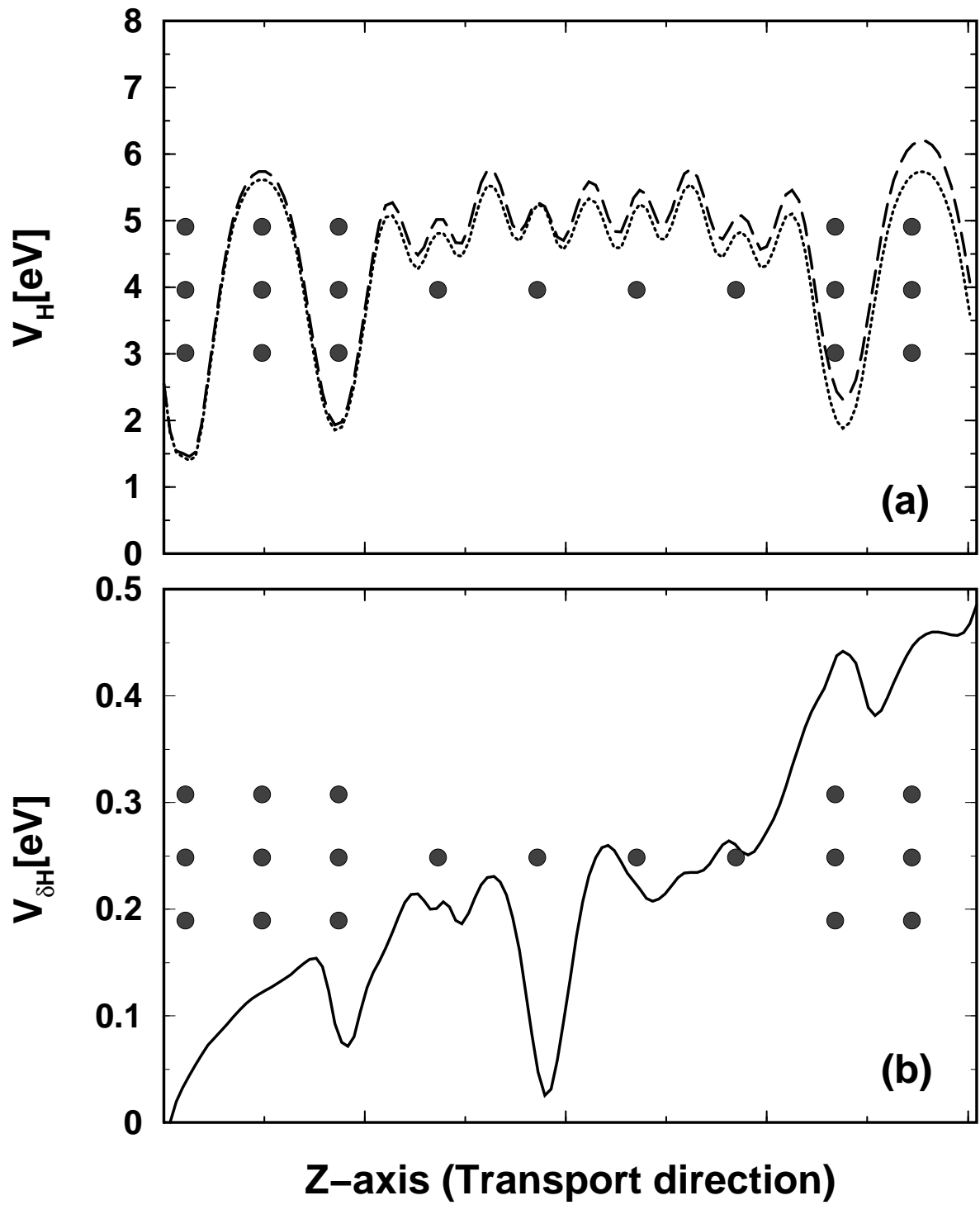
\begin{figure}
\caption{(a) Average Hartree potential ($V_{H}$) across the central 
cross section ($\sim 9\AA^2$) of the ideal contact at
the top site; dotted line at zero bias, and dashed line at $V_b=0.5V$;
and (b) their difference showing that most of the potential 
drops symmetrically at the interfaces.
Black circles correspond to atom positions along the constriction for
guidance}
\label{fig5}
\end{figure}
\begin{figure}
\caption{
Conductance calculated from renormalized transmission due to 
backscattering as a function of bias voltage;
(a) for $T_0=0.99,0.9,0.8$ corresponding to solid, dashed and dotted line,
respectively with interaction parameter $\alpha=1.0$ and (b)
for $T_0=0.9$ and interaction parameter $\alpha=0.1, 0.5, 1.0$
corresponding to solid, dashed and dotted line, respectively. 
}
\label{fig6}
\end{figure}

\begin{thebibliography}{99}
\newcommand{\PRL}{{\em Phys. Rev. Lett }}
\newcommand{\PRB}{{\em Phys. Rev. B }}
\newcommand{\APL}{{\em Appl. Phys. Lett. }}
\newcommand{\sci}{{\em Science }}
\newcommand{\NAT}{{\em Nature } (London) }

\bibitem{agrait1993}
N. Agra\"{\i}t,
J. G. Rodrigo, and S. Vieira,
\PRB {\bf 47}, 12345 (1993).

\bibitem{pascual1993}
J. I. Pascual,
J. M\'endez, J. G\'omez-Herrero, A. M. Bar\'o, and N. Garc\'{\i}a,
and Vu Thien Binh,
\PRL {\bf 71}, 1852 (1993).
\bibitem{olesen1994}
L. Olesen,
E. L$\ae$gsgaard, I. Stensgaard, F. Besenbacher,
J. Schi$\o$tz, P. Stoltze, K. W. Jacobsen, and J. K. N$\o$rskov,
\PRL {\bf72}, 2251 (1994).

\bibitem{pascual1995}
J. I. Pascual,
J. M\'endez, J. G\'omez-Herrero, A. M. Bar\'o, and N. Garc\'{\i}a,
Uzi Landman, W. D. Luedtke, E. N. Bogachek, and H. P. Cheng,
\sci {\bf 267}, 1793 (1995).

\bibitem{branbyge1995}
M. Brandbyge,
J. Schi$\o$tz, M. R. S$\o$rensen, P. Stoltze, K. W. Jacobsen, J. K.
N$\o$rskov,
L. Olesen, E. L$\ae$gsgaard, I. Stensgaard, and F. Besenbacher,
\PRB {\bf 52}, 8499 (1995).

\bibitem{agrait1995}
N. Agra\"{\i}t,
G. Rubio, and S. Vieira,
\PRL {\bf 74}, 3995 (1995).

\bibitem{smith1995}
D. P. E. Smith,
\sci {\bf 169}, 371 (1995).

\bibitem{gai1996}
Zheng Gai,
Yi He, Hongbin Yu, and W. S. Yang,
\PRB {\bf 53}, 1042 (1996).

\bibitem{stalder1996}
A. Stalder and U. D\"urig,
\APL {\bf 68}, 637 (1996).

\bibitem{rubio1996}
G. Rubio,
N. Agra\"{\i}t and S. Vieira,
\PRL {\bf 76}, 2302 (1996).

\bibitem{sirvent1996}
C. Sirvent,
J. G. Rodrigo, S. Vieira, L. Jurczyszyn, N. Mingo, and F. Flores,
\PRB {\bf 53}, 16086 (1996).

\bibitem{costa1997a}
J. L. Costa-Kr\"amer,
\PRB {\bf 55}, R4875 (1997).

\bibitem{costa1997b}
J. L. Costa-Kr\"amer,
N. Garc\"{\i}a, and H. Olin,
\PRL {\bf 78}, 4990 (1997).

\bibitem{untiedt1997}
C. Untiedt,
G. Rubio, S. Vieira, and N. Agra\"{\i}t,
\PRB {\bf 56}, 2154 (1997).

\bibitem{cross1998}
G. Cross,
A. Schirmeisen, A. Stalder,
P. Gr\"utter, M.Tschudy and U. D\"urig,
\PRL {\bf 80}, 4685 (1998).

\bibitem{li1998}
C. Z. Li,
H. Sha, and N. J. Tao,
\PRB {\bf 58}, 6775 (1998).

\bibitem{jian1999}
W. B. Jian,
C. S. Chang, W. Y. Li, and Tien T. Tsong,
\PRB {\bf 59}, 3168 (1999).

\bibitem{krans1993}
J. M. Krans,
C. J. Muller, I. K. Yanson, Th. C. M. Govaert, R. Hesper, and J. M. van
Ruitenbeek,
\PRB {\bf 48}, 14721 (1993).

\bibitem{krans1995}
J. M. Krans, J. M. van Ruitenbeek, V. V. Fisun,
I. K. Yanson, and L. J. de Jongh,
\NAT {\bf 375}, 767 (1995).

\bibitem{zhou1995}
C. Zhou
C. J. Muller, M. R. Deshpande, J. W. Sleight, and M. A. Reed,
\APL {\bf 67}, 1160 (1995).

\bibitem{muller1996}
C. J. Muller,
J. M. Krans,T. N. Todorov, and M. A. Reed,
\PRB {\bf 53}, 1022 (1996).

\bibitem{scheer1997}
E. Scheer,
P. Joyez, D. Esteve, C. Urbina, and M. H. Devoret,
\PRL {\bf 78}, 3535 (1997).

\bibitem{scheer1998}
E. Scheer,
N. Agra\"{\i}t, J. C. Cuevas, A. Levy Yeyati, B. Ludoph,
A. Mart\"{\i}n-Rodero, G. R. Bollinger, J. M. van Ruitenbeek, and C.
Urbina,
\NAT {\bf 394}, 154 (1998).

\bibitem{yanson1998}
A. I. Yanson,
G. R. Bollinger, H. E. van den Brom, N. Agra\"{\i}t, and J. M. van
Ruitenbeek,
\NAT {\bf 395}, 783 (1998).

\bibitem{yasuda1997}
Hiroshi Yasuda and Akira Sakai,
\PRB {\bf 56}, 1069 (1997).

\bibitem{hansen1997}
K. Hansen,
E. L$\ae$gsgaard, I. Stensgaard, and F. Besenbacher,
\PRB {\bf 56}, 2208 (1997).

\bibitem{tekman1991}
E. Tekman and S. Ciraci,
\PRB {\bf 43}, 7145 (1991).

\bibitem{torres1994}
J. A. Torres,
J. I. Pascual, J. J. S\'aenz,
\PRB {\bf 49}, 16581 (1994).

\bibitem{torres1996}
J. A. Torres and J. J. S\'aenz,
\PRL {\bf 77}, 2245 (1996).

\bibitem{garcia1996}
A. Garc\'{\i}a-Mart\'{\i}n,
J. A. Torres, and J. J. S\'aenz,
\PRB {\bf 54}, 13448 (1996).

\bibitem{pascual1997}
J. I. Pascual,
J. A. Torres, and J. J. S\'aenz,
\PRB {\bf 55}, R16029 (1997).

\bibitem{ruitenbeek1997}
J. M. van Ruitenbeek,
M. H. Devoret, D. Esteve, and C. Urbina,
\PRB {\bf 56}, 12566 (1997).

\bibitem{lang1997}
N. D. Lang,
\PRL {\bf 79}, 1357 (1997).

\bibitem{garcia1997}
P. Garc\'{\i}a-Mochales and P. A. Serena,
\PRL {\bf 79}, 2316 (1997).

\bibitem{sanchez1997}
Daniel S\'anchez-Portal,
Carlos Untiedt, Jos\'e M. Soler, Juan J. S\'aenz, and
Nicol\'as Agra\"{\i}t,
\PRL {\bf 79}, 4198 (1997).

\bibitem{wang1997}
C. C. Wang,
J.-L. Mozos, G. Taraschi, Jiang Wang and Hong Guo,
\APL {\bf 71}, 419 (1997).

\bibitem{cuevas1998}
J. C. Cuevas,
A. Levy Yeyati, and A. Mart\'{\i}n-Rodero,
\PRL {\bf 80}, 1066 (1998).

\bibitem{bascones1998}
E. Bascones,
G. G\'omez-Santos, and J. J. S\'aenz,
\PRB {\bf 57}, 2541 (1998).

\bibitem{martin2000}
A. Garc\'{\i}a-Mart\'{\i}n,
M. del Valle, J. J. S\'aenz, J. L. Costa-Kr\"amer and P. A. Serena,
\PRB {\bf 62}, 11139 (2000).

\bibitem{todorov1993}
T. N. Todorov and A. P. Sutton,
\PRL {\bf 70}, 2138 (1993).

\bibitem{bratkovsky1995}
A. M. Bratkovsky,
A. P. Sutton, and T. N. Todorov,
\PRB {\bf 52}, 5036 (1995).

\bibitem{todorov1996}
T. N. Todorov and A. P. Sutton,
{\bf PRB 54}, R14234 (1996);
\bibitem{barnett1997}
R. N. Barnett and Uzi Landman,
\NAT {\bf 387}, 788 (1997).

\bibitem{brandbyge1997a}
Mads Brandbyge,
Karsten W. Jacobsen, and Jens K. N$\o$rskov,
\PRB {\bf 55}, 2637 (1997);

\bibitem{levy1997}
A. Levy Yeyati,
A. Mart\'{\i}n-Rodero, and F. Flores,
\PRB {\bf 56}, 10369 (1997).

\bibitem{mehrez1997a}
H. Mehrez and S. Ciraci,
\PRB {\bf 56}, 12632 (1997).

\bibitem{mehrez1997b}
H. Mehrez,
S. Ciraci, C. Y. Fong and \c{S}. Erko\c{c},
{\em J. of Phys.: Condensed Matter } {\bf 9}, 10843 (1997).

\bibitem{brandbyge1997b}
Mads Brandbyge,
Mads R. S$\o$rensen and Karstn W. Jacobsen,
\PRB {\bf 56}, 14956 (1997).

\bibitem{soresen1998}
Mads R. S$\o$rensen,
Mads Brandbyge, and Karsten W. Jacobsen,
\PRB {\bf 57}, 3283 (1998).

\bibitem{ohnishi1998}
Hideaki Ohnishi,
Yukihito Kondo and Kunio Takayanagi,
\NAT {\bf 395}, 780, (1998).

\bibitem{okamoto1999}
Masakuni Okamoto and Kunio Takayanagi,
\PRB {\bf 60}, 7808 (1999).

\bibitem{choi1999}
Hyoung Joon Choi and Jisoon Ihm,
\PRB {\bf 59}, 2267 (1999).


\bibitem{costa1997}J. L.Costa-Kr\"amer,
N. Garc\'{\i}a, P. Garc\'{\i}a-Mochales, P. A. Serena, M. I. Marqu\'es and
A.Correia,
\PRB {\bf 55}, 5416 (1997).

\bibitem{costa1998}
J. L. Costa-Kr\"amer,N. Garc\'{\i}a, M. Jonson, I. V. Krive, H. Olin, P.A.
Serna and R. I.Shekhter,
Coulomb Blockade effect,Nanoscale Science and Technology, {\bf Vol. 348},
{\em NATO Advanced Study
Institute Series E: Applied Sciences}, Edited by N. Garc\'{\i}a,
M.Nieto-Vesperinas and H. Rohree (Kluwer Academic, Dordrecht, 1998),
p.1.
\label{costa1998}

\bibitem{itakura1999}
Katsuhiro Itakura,
Kenji Yuki, Shu Kurokawa, Hiroshi Yasuda and Akira Sakai,
\PRB {\bf 60}, 11163 (1999).

\bibitem{hansen2000}
K. Hansen,
S. K. Nielsen, M. Brandbyge, E. L$\ae$gsgaard and F. Besenbacher,
\APL {\bf 77}, 708 (2000).

\bibitem{taylor2001}
Jeremy Taylor, Hong Guo, and Jian Wang,
\PRB {\bf 63}, R121104 (2001);

\bibitem{jeremy2}
Jeremy Taylor, Hong Guo, and Jian Wang,
\PRB {\bf 63}, 245407 (2001).

\bibitem{alex2001}
Alex Wlasenko and Peter Gr\"utter
{\em (To be published).}

\bibitem{rice1995}
John A. Rice,
{\em Mathematical Statistics and Data Analysis},
Duxbury Press (Belmont, California) (1995).

\bibitem{eigler}
H. Manoharan, C. Lutz, D. Eigler
{\em (Unpublished).}

\bibitem{hansen2000a}
K. Hansen,
S. K. Nielsen, E. L$\ae$gsgaard, I. Stensgaard and F.
Besenbacher,
{\em Review of Sc. Ins.} {\bf 71}, 1793 (2000).

\bibitem{ordejon1996}
P. Ordej\'on,
E. Artacho, and J. M. Soler,
\PRB {\bf 53}, 10441 (1996).

\bibitem{bachelet1982}
G. B. Bachelet,
D. R. Hamann and M. Schl\"uter,
\PRB {\bf 26}, 4199 (1982).

\bibitem{buttiker1988}
M. B\"uttiker,
{\em IBM J. Res. Dev.} {\bf 32}, 317 (1988).

\bibitem{larade2001}
Brian Larade,
Jeremy Taylor, H. Mehrez, and Hong Guo,
{\em to be published in} \PRB .

\bibitem{agrait2001}
N. Agra\"{\i}t (private communication); In an STM set up
performed in  UHV, initially non-linear I-V characteristics are seen; but
if the tip is retracted and indented into the sample for few times, I-V
characteristics becomes linear. This is clear indication of wearing out
the impurity from the junction.

\bibitem{brandbyge1999}
Mads Brandbyge,
Nobuhiko Kobayashi and Masura Tsukada,
\PRB {\bf 60}, 17064, (1999).
\label{brandbyge1999}

\bibitem{maslov1995}
Dmitrii L. Maslov and Michael Stone,
\PRB {\bf 52}, R5539 (1992).
\label{maslov1995}

\bibitem{datta_book}
S. Datta,
{\em Electronic Transport in Mesescopic Systems},
Cambridge University Press, New York, (1995).

\bibitem{yue1994}
Dongxiao Yue, L. I. Glazman and K. A. Matreev,
\PRB {\bf 49}, 1966 (1994).
\end{thebibliography}
\end{document}